\documentclass{article}
\usepackage{log_2022}

\usepackage{booktabs}            
\usepackage{multirow}            
\usepackage{amsfonts}            
\usepackage{titlesec}
\usepackage{enumerate}
\usepackage{graphicx} 
\usepackage{multirow}
\usepackage{float}

\title{A Multimodal Approach to The Detection and Classification of Skin Diseases}
\author{%
Allen Yang$^{1}$
Edward Yang$^{3}$
$^3$Mission San Jose High School \\
$^3$Department of Computer Science, Yale University \\
Correspondence: \url{edwardyang016@gmail.com}
}

\begin{document}

\section{Title Page}

\textbf{Title:} A Multimodal Approach to The Detection and Classification of Skin Diseases \\
\textbf{Authors:} Allen Yang; Edward Yang \\
\textbf{Affiliation:} Mission San Jose High School, Fremont, CA, USA; Yale University, New Haven, CT, USA \\
\textbf{Correspondence:} Edward Yang. 3075 Olcott St, Santa Clara, CA 95054 +1 (510) 598-5616. Email: edwardyang016@gmail.com. \\
\textbf{ORCID:} E Yang (0000-0003-1911-6364) \\
\textbf{Running title:} A Multimodal Approach to The Detection and Classification of Skin Diseases \\
\textbf{Word Count:} Original Article ($\sim$ 8000) \\
\textbf{Number of Figures\ Tables:} 16 \\
\textbf{Author Contributions:} 
\begin{enumerate}
    \item Conception and design: All authors
    \item Administrative support: E Yang
    \item Provision of study materials or patients: E Yang
    \item Collection and assembly of data: A Yang
    \item Data analysis and interpretation: A Yang
    \item Manuscript writing: All authors
    \item Final approval of manuscript: All authors
\end{enumerate}

\newpage

\section{Abstract}

\textbf{Background:} According to PBS, nearly one-third of Americans lack access to primary care services, and another forty percent delay going to avoid medical costs. As a result, many diseases are left undiagnosed and untreated, even if the disease shows many physical symptoms on the skin. With the rise of AI, self-diagnosis and improved disease recognition have become more promising than ever; in spite of that, existing methods suffer from a lack of large-scale patient databases and outdated methods of study, resulting in studies being limited to only a few diseases or modalities. \\
\textbf{Methods:} This study incorporates readily available and easily accessible patient information via image and text for skin disease classification on a new dataset of 26 skin disease types that includes both skin disease images (37K) and associated patient narratives. Using this dataset, baselines for various image models were established that outperform existing methods. Initially, the Resnet-50 model was only able to achieve an accuracy of 70\% but, after various optimization techniques, the accuracy was improved to 80\%. In addition, this study proposes a novel fine-tuning strategy for sequence classification Large Language Models (LLMs), Chain of Options, which breaks down a complex reasoning task into intermediate steps at training time instead of inference. \\
\textbf{Results:} With Chain of Options and preliminary disease recommendations from the image model, this method achieves state of the art accuracy 91\% in diagnosing patient skin disease given just an image of the afflicted area as well as a patient description of the symptoms (such as itchiness or dizziness). \\
\textbf{Conclusions:} Through this research, an earlier diagnosis of skin diseases can occur, and clinicians can work with deep learning models to give a more accurate diagnosis, improving quality of life and saving lives.\\
\textbf{Keywords:} LLMs, CNNs, Skin Disease Classification, 
\newpage

\section{Highlight Box}
\subsection{Key findings}
\begin{itemize}
    \item Using just skin disease images, the highest accuracy achieved on the new dataset is 80.1\% but by combining it with text data, the accuracy is able to reach 91\% across 26 classes.
\end{itemize}
\subsection{What is known and what is new?}
\begin{itemize}
    \item Doctors use both image and textual/ voice information to diagnosis skin diseases.
    \item Skin disease diagnosis has been explored via image and text separately but never together.
    \item Proposed a new method that utilizes accessible and standard patient information such as skin images and patient narratives to predict the patient’s skin disease.
    \item Introduced a new LLM fine-tuning technique for sequence classification tasks, chain of options, that utilizes chain of thought and precursory information to improve training accuracy.
\end{itemize}
\subsection{What is the implication, and what should change now?}
\begin{itemize}
    \item Demonstrated the potential of the proposed model for detecting multiple skin diseases simultaneously which significantly improves the efficiency and accuracy of skin disease diagnosis.
    \item Proposed a new multi-modal skin disease dataset that is objectively more difficult than existing skin disease benchmarks. 
\end{itemize}
\newpage

\section{Introduction}
\subsection{Background}
According to PBS, nearly one-third of Americans lack access to primary care services, particularly in underserved or remote areas where accessing doctors may be limited, and another forty percent delay going to avoid medical costs. As a result, many diseases are left undiagnosed and untreated, even if the disease shows many physical symptoms. With the rise of Artificial intelligence (AI), self-diagnosis and improved disease recognition have become more promising than ever. AI-driven diagnostic systems can potentially improve the accuracy and speed of disease diagnosis, especially for skin diseases. These tools have shown promising results in the diagnosis of skin diseases, with some studies demonstrating superior performance compared to human dermatologists (1). In the near future, by integrating these tools into mobile applications, individuals can capture images of their skin lesions and receive instant feedback or recommendations for further evaluation. This approach has the advantage of reducing wait times for appointments and providing timely guidance to patients who may not have easy access to doctor services. 

Current state of the art AI models, particularly deep learning algorithms, have demonstrated remarkable capabilities in analyzing skin images and identifying patterns indicative of various skin conditions. These models leverage convolutional neural networks (CNN) to extract features from images and make predictions based on learned patterns.  CNNs consist of multiple layers of convolutional, pooling, and fully connected layers that extract features from input images and classify them into different disease categories. In particular, several vision models including Very Deep Convolutional Networks (VGGNet) (2), ResNet (3), and EfficientNet (4) have been developed, adapted and fine-tuned on various image classification tasks. In fact, EfficientNet has already been attempted for skin lesion classification (5,6). This success in skin disease diagnosis is largely enabled by the development of large datasets, consisting of thousands of annotated images. Additionally, researchers have explored ensemble learning techniques, combining multiple AI models to improve diagnostic performance and robustness.

Despite these promising advancements, AI-driven skin disease diagnosis still faces several challenges and limitations that must be addressed before realizing its full potential in clinical practice. One major challenge is the need for large, high-quality datasets with diverse patient populations and skin disease types. Another challenge is the outdated methods of study, resulting in studies limited to only a few diseases or modalities.  Current neural networks that combine image and text information are only evaluated on 10 diseases or less (7-10). The accuracy from these experiments achieves around 84\% accuracy of prediction (5,6).

\subsection{Rationale and knowledge gap}
In recent years, the application of AI in dermatology has garnered increased attention, offering improvements in the accuracy, efficiency, and accessibility of skin disease diagnosis. From a Google Scholar Search for “A Multimodal Approach to The Detection and Classification of Skin Diseases”, approximately 52,100 results were found. Early methods for skin disease or dermatology image classification were mainly based on traditional machine learning and CNN methods. 

Visual Geometry Group (VGG) (2) is a standard deep Convolutional Neural Network architecture with multiple layers. The “deep” refers to the number of layers in VGG-16 or VGG-19 which have 16 and 19 convolutional layers, respectively. The VGG architecture is the basis of ground-breaking object recognition models. Developed as a deep neural network, VGGNet surpasses baselines on many tasks and datasets beyond ImageNet. Kenneth Thomsen (13) developed a convolutional neural network model to classify clinically relevant selected multiple-lesion skin diseases. A VGG-16 model was trained on 16,543 non-standardized images categorized with ICD-10 codes related to acne, rosacea, psoriasis, eczema, and cutaneous t-cell lymphoma. The best performance was obtained by the VGG-16 model when distinguishing acne from rosacea (sensitivity 85.42\% and specificity 89.53\%). Notably, this model distinguished between the diseases on all three tasks with accuracy above 77\%, indicating a clinically relevant accuracy compared to the reported diagnostic accuracy in dermatology of primary care physicians (48–77\%). Gogineni Saikiran (14) applied VGG to detect the top 10 most common skin diseases with 80\% accuracy. 

Residual Neural Network (ResNet) (3): ResNet introduces shortcut connections, or skip connections, that allow the network to better learn residual mappings, which are the differences between the input and the output of a layer. The ResNet architecture comes in various depths, typically denoted as ResNet-XX, where XX represents the number of layers in the network. For example, ResNet-50 has 50 layers. Many studies have demonstrated the efficacy of ResNet for skin disease detection and classification. For example, Ibrahim Abunadi (15) reported that ResNet-50, when applied to skin disease diagnosis using transfer learning, achieved an accuracy rate of 90\% on the ISIC 2018 dataset and 95.8\% on the PH2 dataset. Many studies have shown that ResNet-based models achieve high accuracy in differentiating between benign and malignant skin lesions, outperforming human dermatologists in certain cases. Additionally, ResNet architectures have been successfully applied to tasks such as melanoma detection, psoriasis diagnosis, and eczema classification, showcasing their versatility and effectiveness in dermatological practice.

EfficientNet (4) studies methods to scale the network architecture across multiple dimensions, including depth, width, and resolution, to achieve better performance without significantly increasing computational cost. After being introduced in 2019, it was immediately applied to skin disease diagnosis. Wu H. (16), using the EfficientNet-B4 model, developed an AI dermatology diagnosis assistant (AIDDA) for Pso, Ecz \& AD and healthy skins (HC). The model was trained on 4,740 clinical images, and performance evaluated using expert-confirmed clinical images grouped into 3 different dermatologist-labeled diagnosis classifications (HC, Pso, Ecz \& AD) with accuracies of 96\%, 89\% and 93\%, respectively. Studies have shown that EfficientNet-based models achieve state-of-the-art accuracy on benchmark datasets for dermatological images, outperforming other CNN architectures while being more computationally efficient.   

Vision Transformer (ViT) (11): A vision model that uses a transformer architecture instead of a CNN for feature extraction. It does this by splitting an input image into several patches, “tokens” and then using an encoder to determine the correlation between tokens. Compared to convolutional methods, ViT is more data hungry and thus will perform better on pretrained or large datasets. Cai (9) proposed a dataset which includes skin disease images and clinical metadata (age, sex, etc.) along with a multimodal Transformer, consisting of two encoders for both image and metadata and one decoder to fuse the multimodal information. A ViT model is used as the backbone to extract image features. As for metadata, they are regarded as labels and a new Soft Label Encoder (SLE) is designed to embed them. Furthermore, in the decoder part, a novel Mutual Attention (MA) block is proposed to better fuse image features and metadata features. To evaluate the model’s effectiveness, extensive experiments have been conducted on the private skin disease dataset and the benchmark dataset ISIC 2018. On the private dataset, the proposed model achieved an accuracy of 0.816, which is better than other popular networks. On the dataset ISIC 2018, the proposed method achieves an accuracy of 0.94 and an AUC of 0.99. 

In conclusion, compared with state-of-the-art methods, the ViT model which combines both images and metadata shows effective performance and advancement in skin disease diagnosis. This study also uses both image and text data for skin disease diagnosis. However, the text data in this case is much more accessible as it represents patient symptoms of each skin disease, patient narratives. This text data could provide key information for diagnosis and help improve diagnostic accuracy. 

\subsection{Objective}
In summary, this method makes the following contributions:
\begin{itemize}
    \item Proposed an end-to-end system that utilizes accessible and standard patient information such as affected skin area images as well as patient narratives to predict the skin disease afflicting the patient
    \item Proposed a new multimodal skin disease dataset and proved that it is objectively more difficult than existing skin disease benchmarks. Evaluated several models and optimization techniques to determine the optimal configuration for the dataset which beats state of the art. 
    \item Introduced a new LLM fine-tuning technique for sequence classification tasks, chain of options, that utilizes chain of thought and precursory information to improve training accuracy and efficiency
    \item Demonstrated the potential of the proposed model for detecting multiple skin diseases simultaneously, which can significantly improve the efficiency and accuracy of skin disease diagnosis.
\end{itemize}

\newpage

\section{Methods}
\subsection{Dataset Development}
\subsubsection{Image Data}
Large datasets comprised of thousands of annotated images are collected from diverse sources, including medical archives, research repositories, and clinical studies. Three such skin disease datasets are available on Kaggle (17, 18, 19). The first dataset consists of 27,153 images classified into ten classes of skin diseases and the second dataset includes 19,600 images classified into 23 types of skin diseases/classes. The first dataset has good distributions of the 10 classes and was used by Hammad (7), who focused on only two classes, which are 1,677 images for eczema and 2,055 images for psoriasis. The third dataset (19), derived from the ISIC competition source website, contains about 23,618 images classified into 8 types of skin diseases.  This study combines the three datasets; however, some images are duplicates or not related to skin diseases. The images were manually checked to see whether they are relevant. Afterwards, hashing was done to remove any duplicate images. The final dataset comprised of 36,995 images across 26 types of skin disease. Unfortunately, class distribution is not perfect because of the large variation in image numbers for different diseases. Some diseases only have around 200 images while others have over 7,900 images. This highlights one of the challenges of getting skin disease images for AI applications which our study overcomes. \textit{Table 1} shows the classes and number of images in each dataset used in this study. The images were divided into 80\% for training and 20\% for validation/testing.

\subsubsection{Text Data from Disease Symptoms}
In addition to images of skin diseases, symptoms of these diseases are another source of information for classifying what ailment a patient has.  This study will add those disease symptoms to help skin disease diagnosis. To do so, the symptoms of skin disease were summarized and aggregated and then turned into patient stories. The symptoms of each skin disease were obtained through Google. \textit{Table 2} shows examples of three diseases and their symptoms summarized from Google. Since the symptoms of each skin disease cannot be directly input into an LLM, this study uses ChatGPT to generate a story to describe a patient’s symptoms based on select symptoms of the disease. The final dataset consists of 10 stories per skin disease with a total of 260 text data. \textit{Table 3} shows an example of 10 stories for the eczema skin disease.

\subsection{Image Classification}
\subsubsection{Initial Development Using Baseline Dataset}
Doctors can discern most skin diseases based solely on images as they generally have unique identifying characteristics. As such, multi-class image classification with neural networks is a popular avenue for classifying skin diseases. Using the first dataset (17), baseline experiments were run with four different neural networks models including VGG, Resnet, Efficientnet, and Vision Transformer to classify 10 different skin diseases (classes).  

Image augmentation techniques have facilitated the generation of synthetic dermatological images, augmenting existing datasets and enhancing the diversity and representativeness of training data. This approach helps mitigate the challenges of dataset imbalance and improves the robustness of AI models to variations in skin types, lighting conditions, and camera quality. Images were initially augmented through several standard techniques such as color jitter, gaussian blur, horizontal/vertical flip, and image resizing. Observing that a majority of “what is important” in the image is in the center of the image as well as the fact that skin diseases are orientation-agnostic, random cropping and random rotation image augmentations were also added.  It is worth noting that, instead of directly random cropping to the desired resolution, a resize was first performed to a size slightly larger than the crop size to retain most of the image proportions. The Resnet-50 model was trained from scratch to determine the optimal data augmentation techniques on an 80: 20 train/ validation split. The first dataset (17) was selected in part because the class distribution was more or less even across the ten classes, simplifying the training process as there would be less worry of overfitting on one particular class. This test establishes a baseline performance that can be used for comparison with the new dataset proposed in this study.

\subsubsection{Applying Base Models to Aggregated New Image Dataset}
Now that baselines for the first dataset (17) have been established, the aggregated new image dataset combines three datasets (17-19) for a total of 26 classes, 16 more than the baseline dataset (10 classes). Due to the new dataset’s class imbalance, reaching model convergence would be significantly harder than on the original dataset. To initially combat this issue, weighted random sampling was introduced which, instead of iterating through the whole train dataset per epoch, a weighted random sample was selected with the chance of an image from a particular class being selected as the inverse of the frequency. In this way, classes with less data are seen more often, reducing the effect of class imbalance. The same models from the baseline study are re-evaluated on this new dataset and the performance was compared.  

\subsubsection{Transfer Learning Comparison}
To further combat class imbalance, transfer learning was used. Transfer learning occurs when 100\% of the feature extractor is frozen and only the classifier is trained on a new dataset. In this case, the feature extractor is initialized to ImageNet weights. By using transfer learning, the model doesn’t need to learn a feature extractor and classifier simultaneously. This is especially useful in an imbalanced dataset as a trainable feature extractor would only learn the relevant features in the most common classes. Instead, a pretrained feature extractor would already understand what features are important in an image and enable the classifier to decide what disease these extracted features are related to.

\subsubsection{Finetuning of Pretrained Resnet Model}
However, directly using ImageNet-trained feature extractor weights has some issues. This is because the data distribution for the pretrained dataset, ImageNet, is significantly different from the current use case. While ImageNet contains pictures of dogs, cats, and other easily identifiable objects, the current dataset contains only skin disease images which have minute differences that correspond to vastly different diseases. As a result, what may be a distinguishing feature between a dog and a cat can’t be used to distinguish between melanoma and basal cell carcinoma. To address this, various percentages of the model layers were unfrozen (as compared to completely frozen in previous experiment) and allowed to be trained in order to determine the optimal percent at which the model would achieve the highest accuracy. By doing so, the model would be able to utilize the high level, class- agnostic, features learned through pretraining such as the outlines of shapes which are commonly found in the initial layers of the feature extractor. The model would then only need to adjust the later layers of the feature extractor which are responsible for low level features that are specific to the task at hand. 

\subsubsection{Adjusting Image Resolution}
To further improve performance, image resolution was also adjusted. Since higher resolution images offer more visual cues of what the skin disease is, various models at 75\% pretrained were evaluated. Beginning with the initial resolution of 224 x 224 pixels, the images were enlarged up to 528 x 528. At the initial 224 x 224 resolution used in previous works, the original image would need to be shrunk by approximately 80\%, significantly reducing the number of features available. By increasing the resolution, the resulting image would have more features available for the model to discern using which might detract from performance as well since many of these features could be extraneous and not related to a disease. This study aims to determine the optimal resolution for images on this new dataset.

\subsection{Text Classification}
\subsubsection{Baseline Development}
In conventional patient- doctor interactions, not only is the picture of the afflicted skin region available to the doctor but also a patient narrative that describes what symptoms the patient is experiencing. This information is important as some diseases may look the same from a physical inspection but cause the patient to exhibit different symptoms which are not apparent from an image. Therefore, based on the symptoms, the doctor should also be able to determine what type of skin disease the patient has. In the context of machine learning, this could be considered a multi class sequence classification task which large language models (LLM) excel at. Thus, three state of the art large language models (Llama-7B (20), Falcon-7B (21), Mistral-7B (22)) were fine- tuned on this dataset. This fine- tuning step is necessary in order for the model to understand what a proper sentence entails (from pretraining) as well as what specific keywords are associated with a particular disease. Due to computational constraints associated with such large models, LLMs larger than 7B were unable to be run and the three selected LLMs were finetuned using Low Rank Adaptation (LoRA). By using LoRA, only a small, low rank, subset of the model’s parameters (7 - 10\%) need to be fine-tuned which enables the models to fit in consumer-grade GPUs.  The dataset itself is split 70: 30 train/ validation due to its small size. 

\subsubsection{Choice/ Options Mapping}
Along with the patient narrative text block input for sequence classification, additional supporting information was also added. As LLMs are trained on millions of corpuses that describe hundreds of diseases, it may be difficult for them to associate the presented symptoms to only the subset used in this study. As such, it makes sense to restrict the potential set of diseases the LLM must choose from to the subset by presenting the LLM, in its prompt, the available disease options. Initially, an explicit options mapping was concatenated at the end of each patient narrative text block which mapped each disease to its associated value such as Dermatofibroma to “1” and Lichen Planus to “7”. By explicitly adding this mapping, the LLM would only need to know the disease associated with the narrative as the mapping is given instead of learning both the disease and then mapping. Additionally, an implicit options list was also explored which concatenated only the possible diseases without their associated values as shown in \textit{Table 8}.

\subsubsection{Chain of Options}
Inspired by Chain of Thought prompting, a popular prompting technique that breaks down a complex reasoning task into intermediate steps for the LLM to take, Chain of Options breaks down the task during fine tuning instead of inference. This allows for more control in the reasoning process as the LLM’s performance could be evaluated during intermediate steps instead of only at the end. At a high level, Chain of Options incrementally removes k options from the options list based on which diseases are the most unlikely. Gradually, the options list would decrease until there are less than k diseases left at which point the top prediction is considered the final prediction. This “narrowing down” process reframes the task from needing to pick the most likely disease out of 26 options, which could be difficult if many diseases have similar symptoms, into a task to simply determine which k diseases are the least likely. This is a simpler reframe as picking the least likely diseases could be from just a single symptom mismatch. During training, random combinations of diseases of varying lengths are concatenated in the same way as the options list was. This randomness simulates the various instances of the chain that could potentially propagate through while also adding some noise to the input to reduce overfitting.

\subsection{Combining Image and Text Classification}
Along with providing an options list, the output from the image model is also added to the narrative, imitating an “expert” (LLM) having the final say in what disease a patient has, given initial recommendations (image model). To imitate this process, the top N most likely predictions from the image model would be used as an initial prediction which are then included alongside the patient narrative as input into the LLM. \textit{Figure 1} describes the architecture of the model for skin diseases classification in this study. Conceptually, this would further decrease the number of potential diseases the LLM must choose from to the top N. In practice, this is not the case as the image model is not entirely accurate even when the top 5 predicted diseases are considered. As a result, the LLM must not completely assume that the correct disease is within the recommended diseases although it is very likely it is. To imitate this in training, instead of always including the correct disease in the recommendation, there is a chance that the recommendation doesn’t actually include it, with the chance being the same as the prediction accuracy of the image model.

\newpage

\section{Results}
The experiments were done on a home computer with an RTX 4090 GPU, Intel Core i9-13900K CPU, and 32 GB DDR5 memory. \textit{Table 4} shows the effect of adding various image augmentation techniques during training for various image models including Resnet, Efficientnet, and ViT. The baseline is the top results reported for that dataset according to Kaggle, \textit{Table 5} shows that the new proposed dataset is significantly more difficult as the model performance decreased across the board. \textit{Table 6} shows that transfer learning is able to improve the results on some models but not others. \textit{Figure 2} shows the effect of the proportion of the pre-trained model being frozen versus accuracy as well as the epochs until convergence. \textit{Table 7} shows the effect of increasing image resolution on the accuracy across select models. \textit{Figure 3} shows the confusion matrix for the optimal vision model on the new image dataset. \textit{Table 8} shows the baseline accuracies of three LLMs with various augmentation techniques. \textit{Table 9} shows the accuracies with various prediction augmentations included. \textit{Table 10} shows the accuracies after adding Chain of Options, \textit{Table 11} shows the final results after combining the optimal configuration for the vision and large language models.
\newpage

\section{Discussion}
\subsection{Image Augmentation and Model Comparisons}
Baselines for multiple models were established on the 1st dataset. Several image augmentation techniques were used, beginning with standard ones such as color jitter, gaussian blur, horizontal/ vertical flip and image resizing. These augmentations improved validation accuracy from 70\% to 76\% as shown in \textit{Table 4}. Observing that a majority of “what is important” in the image is in the center of the image as well as the fact that skin diseases are orientation- agnostic, random cropping and random rotation image augmentations were added, improving performance to 82\% on the first dataset. Using this augmentation configuration, baselines for Efficientnet- B3, ViT, and VGG- 11 were also run with similar validation accuracies in \textit{Table 4}. These validation accuracies by Efficientnet- B3, ViT, Resnet-50, and VGG- 11 were 81.6\%, 82.2\%, 81.7\% and 78.1\% respectively, which were in line with the top performing models within Kaggle.
\subsection{Classification on Aggregated New Image Dataset}
\subsubsection{Applying Baselines to New Dataset}
Four base models were evaluated on the aggregated new image dataset which contains 36995 images spanning 26 classes. Their results are compared to the first dataset (10 classes) in \textit{Table 5}.  Across the board, model performance decreased with VGG- 11 and ViT failing to converge and Resnet- 50 and Efficientnet- B3 accuracy dropping by at least 10\%, proving that this dataset, due to its class imbalance, was significantly harder than the original dataset. 
\subsubsection{Optimized Models for New Dataset}
\noindent\textbf{Transfer Learning}

 Since VGG-11 and ViT failed to converge when trained from scratch on this dataset, the transfer learning paradigm was used. Using this method, the training speed of the models significantly increased with most models reaching convergence in around 4 hours as opposed to 8 hours in the from scratch case. ViT was able to converge although VGG-11 and 16 still failed to do so. This makes sense as ViT isn’t able to leverage the inductive bias that convolutional neural networks (VGG, Resnet, and EfficientNet) have inherently baked in since ViT uses transformer layers as opposed to convolutional layers. Therefore, when the dataset is not only small but also imbalanced, ViT struggles to learn as it must learn inductive biases toward images as well as the classes of the images simultaneously. By using transfer learning, ViT has already learned the inductive biases toward images and would only need to learn the image classes for a new dataset. VGG, on the other hand, struggles from the “vanishing gradient” problem. Due to the depth of the network, gradient updates to initial layers of the network become very small, significantly increasing the training time. However, without such a large network, the intricacies of the dataset are more difficult to learn as the model isn’t able to fit as complex of functions. 
 
\textit{Table 6} compares the epochs and accuracies between not pretrained and pre-trained with 100\% feature extractor frozen (transfer learning) models.  Directly using pre-trained feature extractors lead to suboptimal performance for models that are able to initially converge for both Resnet and Efficientnet models. However, pre-training helps improve performance for all other models (VGG, ViT).

\noindent\textbf{Fine-Tuning of Pre-trained Resnet Model}

For this task, the best performing model (Resnet-50) across both not pre-trained and 100\% pre-trained was used and it was determined that 75\% frozen parameters is optimal as shown in \textit{Figure 2}. As the percent of frozen parameters increased, the number of epochs it took for the model to converge decreased. Similarly, as the percent of frozen parameters increased, so did the accuracy up to a certain point. Exceeding 75\% frozen, the accuracy begins to decrease. Conceptually, this amount is the perfect balance between the high level, class- agnostic, features learned through pre-training and the low level, task- specific, features learned during fine-tuning. 

\noindent\textbf{Fine-Tuning of Image Resolution}

The top performing model was found to be Resnet50 with images at 300 x 300 resolution. This model achieved 80.1\% top-1 accuracy which is 8\% higher than the baseline model. Further analysis shows that this model achieves 92.1\% and 95.2\% top-3 and top-5 accuracy, respectively. Increasing the resolution past 300 x 300, accuracy improvement saturates as the model accuracy is about the same between 300 x 300 and 528 x 528 resolution.

In conclusion, after optimizing the Resnet-50 model through image augmentation, fine-tuning, image resolution adjustment, the model with image resolution of 300x300 and 75\% pre-trained on the new image dataset (36995 Images and 26 classes) achieved 80.1\%, 92.1\% and 95.2\% for top-1, top-3 and top-5 accuracy, respectively as shown in \textit{Figure 3}.

\subsubsection{Large Language Models (LLM) For Text Data}

\noindent\textbf{Choice/ Options Mapping} The implicit options list outperformed the explicit map, proving that learning the mapping between disease and label is quite simple and that the mapping adds too much duplicate information which would actually hurt LLM performance. Furthermore, it demonstrates that LLMs perform better on multi- class sequence classification tasks when all the possible options are provided. By specifying which diseases, the LLM should consider through an explicit/ implicit options map, LLM performance is improved. In most cases, the implicit options map outperforms the explicit one with highest accuracy (89.7\%) being achieved by Llama -7B  and implicit options map.
\linebreak

\noindent\textbf{Prediction Concatenation} 
In order to utilize the predictions from the image model, they were concatenated onto the patient narrative in a similar way as choice/ options mapping. With this addition to the narrative, model accuracy increased by 20\% from the baseline model for all three LLMs as shown in \textit{Table 9}. Adding only the top-1 prediction had the highest improvement while top-3 and top-5 had slightly less. This is interesting because, as the number of provided predictions increased, the accuracy of the LLM decreased despite the likelihood of the correct class being in the predictions increasing as well. The same phenomenon occurs when we combine predictions and choice/ options mapping: although performance is better than just choice/ options mapping, performance decreases as the number of predictions increases. 
\linebreak

\noindent\textbf{Chain of Options}
To mitigate the issue of accuracy decreasing with more provided predictions, a novel training strategy was introduced called Chain of Options. By using Chain of Options, LLM performance for higher numbers of recommendations improved across all LLMs while maintaining at least a 90\% accuracy as shown in \textit{Table 10}.

\subsubsection{Combining Vision Model and LLM for Skin Disease Diagnosis}

\hspace{\parindent} Using resnet-50, image alone is only able to achieve ~80\% accuracy on the new dataset as discussed in the previous section. However, by combining both the top performing image model resnet-50 and the best LLM (Llama-7B), a final skin disease classification accuracy of 91\% was achieved with the image model recommending the top 5 candidate diseases and the LLM adding these candidates to the input along with chain of options fine-tuning. This improved the overall accuracy as seen in \textit{Table 11}.

In summary, the Resnet-50 model was applied to the new image dataset and the top-1 validation accuracy was 70\% for 26 types of skin disease diagnosis. After further optimizing the model through various means such as fine-tuning, top-1 accuracy was improved to 80\%. The best LLM, by itself, was able to achieve 76\% accuracy. By combining the top performing image model with the top LLM, a final accuracy of 91\% was achieved in \textit{Figure 4}. This result outperforms human dermatologists in certain cases for over 26 types of skin disease diagnosis.

\newpage

\section{Conclusions}
This work presents a novel method for classifying skin diseases using AI vision and large language models. A new multimodal dataset was proposed containing both skin disease images and associated patient narratives for 26 skin disease types. Baselines for several state-of-the-art image models including  VGG, Resnet, Efficientnet, and ViT were established which outperform existing methods. It was found that the Resnet-50 model with 75\% of its parameters frozen performs best on this dataset with a final top-1 accuracy of 80.1\%. This study also proposes a novel fine-tuning strategy for sequence classification LLMs, Chain of Options, which breaks down a complex reasoning task into intermediate steps at training time instead of inference. With Chain of Options and preliminary disease recommendations from the image model, this method achieves 91.2\% accuracy in diagnosing patient skin disease given just an image of the afflicted area as well as a patient description of the symptoms (such as itchiness or dizziness) using Llama-7B fine-tuned with LoRA.  As such, this study demonstrated the potential of the proposed models for detecting multiple skin diseases simultaneously, which can significantly improve the efficiency and accuracy of skin disease diagnosis. 
Future work includes: 
\begin{itemize}
    \item Expanding dataset to include more skin disease types
    \item Implementing Retrieval- Augmented Generation (RAG) for on-demand disease fact retrieval using external knowledge bases
    \item Developing an application for real- world use/ testing
    \item Integrating with smartphones, individuals can capture images of their skin lesions and receive instant feedback or recommendations.
\end{itemize}

\newpage

\section{Acknowledgments}
This project was self-funded.

\section{Footnote}
\textbf{Data Sharing Statement:} Please see upload\\
\textbf{Peer Review File:} TBA\\
\textbf{Conflicts of Interest:} The authors have no conflicts of interests to declare. Please see upload\\
\textbf{Ethical Statement:} The authors are accountable for all aspects of the work in ensuring that questions related to the accuracy or integrity of any part of the work are appropriately investigated and resolved.

\section{References}
\begin{enumerate}
\item Esteva A, Kuprel B, Novoa RA, Ko J, Swetter SM, Blau HM, et al. Dermatologist-level Classification of Skin Cancer with Deep Neural Networks. Nature [Internet]. 2017 Jan 25;542(7639):115–8. Available from: https://www.nature.com/articles/nature21056
\item Boesch G. VGG Very Deep Convolutional Networks (VGGNet) - What you need to know [Internet]. viso.ai. 2021. Available from: https://viso.ai/deep-learning/vgg-very-deep-convolutional-networks/
\item ResNet: The Basics and 3 ResNet Extensions [Internet]. Datagen. Available from: https://datagen.tech/guides/computer-vision/resnet/
\item Tan M, Le Q. EfficientNet: Rethinking Model Scaling for Convolutional Neural Networks [Internet]. 2019. Available from: https://arxiv.org/pdf/1905.11946.pdf
\item Liopyris K, Gregoriou S, Dias J, Stratigos AJ. Artificial Intelligence in Dermatology: Challenges and Perspectives. Dermatology and Therapy. 2022 Oct 28;12(12).
\item Li Z, Koban KC, Schenck TL, Giunta RE, Li Q, Sun Y. Artificial Intelligence in Dermatology Image Analysis: Current Developments and Future Trends. Journal of Clinical Medicine. 2022 Nov 18;11(22):6826.
\item Hammad M, Paweł Pławiak, ElAffendi M, Abd AA, Abdel A. Enhanced Deep Learning Approach for Accurate Eczema and Psoriasis Skin Detection. Sensors. 2023 Aug 21;23(16):7295–5.
\item Jain A, Way D, Gupta V, Gao Y, de Oliveira Marinho G, Hartford J, et al. Development and Assessment of an Artificial Intelligence–Based Tool for Skin Condition Diagnosis by Primary Care Physicians and Nurse Practitioners in Teledermatology Practices. JAMA Network Open [Internet]. 2021 Apr 28;4(4):e217249–9. Available from: https://jamanetwork.com/journals/jamanetworkopen/fullarticle/2779250
\item Cai G, Zhu Y, Wu Y, Jiang X, Ye J, Yang D. A multimodal transformer to fuse images and metadata for skin disease classification. The Visual Computer. 2022 May 5;
\item Escalé-Besa A, Yélamos O, Vidal-Alaball J, Fuster-Casanovas A, Miró Catalina Q, Börve A, et al. Exploring the potential of artificial intelligence in improving skin lesion diagnosis in primary care. Scientific Reports [Internet]. 2023 Mar 15;13(1):4293. Available from: https://www.nature.com/articles/s41598-023-31340-1
\item Dosovitskiy A, Beyer L, Kolesnikov A, Weissenborn D, Zhai X, Unterthiner T, et al. AN IMAGE IS WORTH 16X16 WORDS: TRANSFORMERS FOR IMAGE RECOGNITION AT SCALE [Internet]. 2021 Jun. Available from: https://arxiv.org/pdf/2010.11929.pdf
\item OpenAI. ChatGPT [Internet]. chat.openai.com. OpenAI; 2024. Available from: https://chat.openai.com/chat
\item Thomsen K, Christensen AL, Iversen L, Lomholt HB, Winther O. Deep Learning for Diagnostic Binary Classification of Multiple-Lesion Skin Diseases. Frontiers in Medicine. 2020 Sep 22;7.
\item Gogineni Saikiran, Narayana GS, Dhanrajnath Porika, Kumar GV. Clinical Skin Disease Detection and Classification: Ensembled VGG [Internet]. 2020. p. 827–47. Available from: https://www.researchgate.net/publication/345434604\_Clinical\_
Skin\_Disease\_Detection\_and\_Classification\_Ensembled\_VGG
\item Abunadi I, Senan EM. Deep Learning and Machine Learning Techniques of Diagnosis Dermoscopy Images for Early Detection of Skin Diseases. Electronics. 2021 Dec 18;10(24):3158.
\item Wu H, Yin H, Chen H, Sun M, Liu X, Yu Y, et al. A deep learning, image based approach for automated diagnosis for inflammatory skin diseases. Annals of Translational Medicine. 2020 May;8(9):581–1.
\item Skin diseases image dataset [Internet]. www.kaggle.com. Available from: https://www.kaggle.com/datasets/ismailpromus/skin-diseases-image-dataset/data
\item Dermnet [Internet]. www.kaggle.com. Available from: https://www.kaggle.com/datasets/shubhamgoel27/dermnet/data
\item ISIC - 2019 [Internet]. www.kaggle.com. Available from: https://www.kaggle.com/datasets/bhanuprasanna/isic-2019
\item Touvron H, Lavril T, Izacard G, Martinet X, Lachaux MA, Lacroix T, et al. LLaMA: Open and Efficient Foundation Language Models [Internet]. 2023 Feb. Available from: https://arxiv.org/pdf/2302.13971.pdf
\item AI M. Mistral AI | Open source models [Internet]. mistral.ai. Available from: https://mistral.ai/
\item Falcon LLM [Internet]. 2024 [cited 2024 Aug 9]. Available from: https://falconllm.tii.ae/falcon-models.html

\end{enumerate}

\section{Tables}
\begin{table}[H]
\caption{Class Statistics for Skin Disease Image Datasets}
\label{tab:Disease_Dataset}
\resizebox{\columnwidth}{!}{\begin{tabular}{|l|lll|l|}
\hline
\multirow{2}{*}{\textbf{Class Name (Skin Diseases)}} & \multicolumn{3}{l|}{\textbf{Number of Images}}                                    & \multirow{2}{*}{\textbf{Final Dataset}} \\ \cline{2-4}
                                                     & \multicolumn{1}{l|}{1st Dataset} & \multicolumn{1}{l|}{2nd Dataset} & 3rd Dataset &                                         \\ \hline
Acne and Rosacea                                     & \multicolumn{1}{l|}{}            & \multicolumn{1}{l|}{1152}        &             & 858                                     \\ \hline
Atopic Dermatitis                                    & \multicolumn{1}{l|}{1257}            & \multicolumn{1}{l|}{612}            &             &         1210                                \\ \hline
Basal Cell Carcinoma (BCC)                           & \multicolumn{1}{l|}{3323}            & \multicolumn{1}{l|}{1437}            &      3323       &       4709                                  \\ \hline
Benign Keratosis-like Lesions (BKL)          & \multicolumn{1}{l|}{2079}            & \multicolumn{1}{l|}{}            &      2624       &                    2065                     \\ \hline
Bullous Disease                         & \multicolumn{1}{l|}{}            & \multicolumn{1}{l|}{561}            &             &                   528                      \\ \hline
Cellulitis Impetigo and other Bacterial Infections     & \multicolumn{1}{l|}{}            & \multicolumn{1}{l|}{361}            &             &       352                                  \\ \hline
Eczema                              & \multicolumn{1}{l|}{1677}            & \multicolumn{1}{l|}{1544}            &             &    1553                                     \\ \hline
Exanthems and Drug Eruptions                  & \multicolumn{1}{l|}{}            & \multicolumn{1}{l|}{505}            &             &           467                              \\ \hline
Hair Loss Alopecia and other Hair Diseases     & \multicolumn{1}{l|}{}            & \multicolumn{1}{l|}{299}            &             &         282                                \\ \hline
Light Diseases and Disorders of Pigmentation         & \multicolumn{1}{l|}{}            & \multicolumn{1}{l|}{711}            &             &      676                                   \\ \hline
Lupus and other Connective Tissue diseases       & \multicolumn{1}{l|}{}            & \multicolumn{1}{l|}{525}            &             &           511                              \\ \hline
Melanocytic Nevi (NV)                            & \multicolumn{1}{l|}{7970}            & \multicolumn{1}{l|}{}            &     12896        &     7967                                    \\ \hline
Melanoma Skin Cancer Nevi and Moles   & \multicolumn{1}{l|}{3140}            & \multicolumn{1}{l|}{579}            &      4522       &            3698                             \\ \hline
Nail Fungus and other Nail Disease   & \multicolumn{1}{l|}{}            & \multicolumn{1}{l|}{1301}            &             &                   1163                      \\ \hline
Poison Ivy and other Contact Dermatitis                     & \multicolumn{1}{l|}{}            & \multicolumn{1}{l|}{325}            &             &       308                                  \\ \hline
Psoriasis pictures Lichen Planus and related diseases                   & \multicolumn{1}{l|}{2055}            & \multicolumn{1}{l|}{1757}            &             &  1791                                       \\ \hline
Scabies Lyme Disease and other Infestations and Bites                  & \multicolumn{1}{l|}{}            & \multicolumn{1}{l|}{539}            &             &     479                                    \\ \hline
Seborrheic Keratoses and other Benign Tumors         & \multicolumn{1}{l|}{1847}            & \multicolumn{1}{l|}{1714}            &             &          1802                               \\ \hline
Systemic Disease                              & \multicolumn{1}{l|}{}            & \multicolumn{1}{l|}{758}            &             &            698                             \\ \hline
Tinea Ringworm Candidiasis and other Fungal Infections       & \multicolumn{1}{l|}{1702}            & \multicolumn{1}{l|}{1625}            &             &     1546                                    \\ \hline
Urticaria Hives                & \multicolumn{1}{l|}{}            & \multicolumn{1}{l|}{265}            &             &              261                           \\ \hline
Vascular Tumors                                & \multicolumn{1}{l|}{}            & \multicolumn{1}{l|}{603}            &      253       &             845                            \\ \hline
Vasculitis                                              & \multicolumn{1}{l|}{}            & \multicolumn{1}{l|}{521}            &             &      510                                   \\ \hline
Warts Molluscum and other Viral Infections                                   & \multicolumn{1}{l|}{2103}            & \multicolumn{1}{l|}{1358}            &             &           1849                              \\ \hline
Squamous cell carcinoma                       & \multicolumn{1}{l|}{}            & \multicolumn{1}{l|}{}            &     628        &          628                               \\ \hline
Dermatofibroma                                          & \multicolumn{1}{l|}{}            & \multicolumn{1}{l|}{}            &    239         &     239                                    \\ \hline
\textbf{Total Diseases (Classes)}                    & \multicolumn{1}{l|}{\textbf{10}}   & \multicolumn{1}{l|}{\textbf{22}}   & \textbf{7}   & \textbf{26}                               \\ \hline
\textbf{Total Images}                                & \multicolumn{1}{l|}{\textbf{27153}}   & \multicolumn{1}{l|}{\textbf{19052}}   & \textbf{23618}   & \textbf{36995}                               \\ \hline
\end{tabular}}
\end{table}

\begin{table}[H]
\caption{Examples of three diseases and their symptoms}
\label{tab:Disease_Example}
\resizebox{\columnwidth}{!}{\begin{tabular}{|p{2cm}|p{6cm}|p{6cm}|}
\hline
\textbf{Disease Name} & \textbf{Symptoms From Google} & \textbf{Symptoms from ChatGPT} \\ \hline
    Basal Cell Carcinoma (BCC)      &          A shiny, skin-colored bump that's translucent, meaning you can see a bit through the surface. The bump can look pearly white or pink on white skin. On brown and Black skin, the bump often looks brown or glossy black. Tiny blood vessels might be visible, though they may be difficult to see on brown and Black skin. The bump may bleed and scab over. A brown, black or blue lesion — or a lesion with dark spots — with a slightly raised, translucent border. A flat, scaly patch with a raised edge. Over time, these patches can grow quite large. A white, waxy, scar-like lesion without a clearly defined border.

                     &               Basal Cell Carcinoma (BCC) is the most common type of skin cancer, usually caused by exposure to ultraviolet (UV) radiation from sunlight. It typically appears as a pearly or waxy bump on the skin, often with visible blood vessels, or as a flat, scaly, reddish patch. BCC tends to grow slowly and rarely metastasizes, but it can cause disfigurement if left untreated. Actinic Keratosis (AK) is a precancerous skin lesion caused by long-term exposure to UV radiation. It appears as rough, scaly patches on the skin, commonly found on sun-exposed areas such as the face, scalp, and hands. While AK itself is not cancerous, it has the potential to develop into squamous cell carcinoma if left untreated. 

                 \\ \hline
    Benign Keratosis-like Lesions (BKL)       &            A round or oval-shaped waxy or rough bump, typically on the face, chest, a shoulder or the back
A flat growth or a slightly raised bump with a scaly surface, with a characteristic "pasted on" look
Varied size, from very small to more than 1 inch (2.5 centimeters) across; Varied number, ranging from a single growth to multiple growths; Very small growths clustered around the eyes or elsewhere on the face, sometimes called flesh moles or dermatosis papulose nigra, common on Black or brown skin; Varied in color, ranging from light tan to brown or black; Itchiness

                   &            BKL encompass a range of skin growths, including seborrheic keratoses and other benign tumors. Seborrheic keratoses are common non-cancerous growths that appear as wart-like or stuck-on lesions on the skin's surface. They can vary in color, size, and texture, often presenting as brown, black, or tan growths with a waxy, scaly, or rough surface. While seborrheic keratoses are typically harmless, they may be cosmetically bothersome or cause itching. Other benign tumors, such as dermatofibromas or skin tags, may also be categorized under BKL. These growths are generally benign and do not pose a significant health risk, but they may be removed for cosmetic reasons or if they cause discomfort.

                    \\ \hline
    Eczema      &      Dry, cracked skin; Itchiness (pruritus); Rash on swollen skin that varies in color depending on your skin color; Small, raised bumps, on brown or Black skin; Oozing and crusting; Thickened skin; Darkening of the skin around the eyes
Raw, sensitive skin from scratching                        &                   Eczema, also known as atopic dermatitis, is a chronic inflammatory skin condition characterized by dry, itchy, and red patches of skin. It can occur at any age but is most common in children. The exact cause of eczema is not fully understood, but it is believed to involve a combination of genetic and environmental factors. Triggers for eczema flare-ups can include irritants like soaps and detergents, allergens like dust mites or pet dander, changes in weather or humidity, and stress. 

             \\ \hline
\end{tabular}}
\end{table}

\begin{table}[H]
\caption{Examples of Eczema Text Data from Symptoms Generated by ChatGPT}
\label{tab:Example_Data}
\resizebox{\columnwidth}{!}{\begin{tabular}{|p{2cm}|p{3cm}|p{3cm}|p{9cm}|}
\hline
\textbf{Eczema Disease} & \textbf{Keywords} & \textbf{Prompt (GPT Question)} & \textbf{Story (GPT Response)} \\ \hline
  Example 1     &          Dry, cracked skin, Itchiness, Small, raised bumps, on brown or Black skin, Oozing and crusting, Thickened skin, sensitive skin from scratching

         &            Pretending you are a patient, please construct a one paragraph patient narrative using these symptoms: "{keywords}"                    &            I've been dealing with a myriad of skin issues lately that have become increasingly bothersome. My skin feels incredibly dry and cracked, causing constant discomfort. The itching is relentless, and I can't resist scratching, leading to raw and sensitive areas. What's most concerning is the rash on my swollen skin, which changes color based on my skin tone, particularly noticeable on my brown skin. Small, raised bumps have appeared, oozing and crusting in some areas, while the skin has thickened in others. To add to my worries, the skin around my eyes has darkened. It's been a perplexing and distressing experience, and I'm eager to find out what's causing these symptoms and how to alleviate them.

                   \\ \hline
       2                 &       Dry, cracked skin, Itchiness (pruritus), Rash on swollen skin that varies in color depending on your skin color

            &          Pretending you are a patient, please construct a one paragraph patient narrative using these symptoms: "{keywords}"

                      &          I've been grappling with a series of unsettling skin issues that are really affecting my daily life. My skin feels incredibly dry and cracked, and the constant itchiness has become maddening. The appearance of a rash on my swollen skin is particularly troubling, and what's strange is that it varies in color depending on my skin tone. It's not just uncomfortable; it's also visually distressing. I can't seem to find relief, and I'm anxious to understand what's causing these symptoms and how I can get some relief for my skin.

                     \\ \hline
         3               &      Rash on swollen skin, Oozing and crusting and Thickened skin

             &             I would like to talk to a doctor and would like to describe my symptoms: Rash on swollen skin, Oozing and crusting and Thickened skin. please construct a one paragraph patient narrative using those symptoms

                   &                   I've been experiencing a rash that appears on swollen skin, accompanied by oozing and crusting. The affected area also seems to have thickened skin. The rash is red and raised, and the surrounding skin appears puffy. There's discharge coming from the rash, with a yellowish color and crust forming over time. The skin feels tougher than usual, with a rough texture. These symptoms started recently and haven't improved with home remedies. I'm concerned about what might be causing this and would appreciate your insight and guidance.

            \\ \hline
          4              &     cracked skin
Itchiness (pruritus)
Raw, sensitive skin from scratching

              &              Pretending you go to hospital with your daughter to see the skin doctor, please help write a paragraph to describe her symptoms: cracked skin, Itchiness (pruritus) and Raw, sensitive skin from scratching

                 &              My daughter has been experiencing cracked skin, relentless itchiness (pruritus), and her skin has become raw and sensitive from scratching when she was infancy. The cracks in her skin are visible, especially on her arms and legs, and they seem to worsen with movement. She constantly complains of itching, which she tries to relieve by scratching, leading to further irritation and discomfort. Her skin has become red and inflamed in areas where she scratches the most, and it's evident that she's in significant discomfort. We're hoping to find relief and a solution to alleviate her symptoms.

                 \\ \hline
           5             &    Oozing and crusting, Dry skin, pruritus

               &               Pretending you are a patient, please construct a one paragraph patient narrative using these symptoms: Oozing and crusting, Dry skin, pruritus

                 &                 I've been experiencing oozing and crusting on my skin, particularly in areas where it's excessively dry. Alongside this, I've been dealing with persistent pruritus, causing intense itching that's difficult to relieve. The dryness of my skin seems to exacerbate the itching, leading to further irritation and discomfort. The oozing and crusting are especially concerning, as they indicate a deeper issue with my skin's health. I'm seeking help to alleviate the discomfort and address the underlying cause of these symptoms
              \\ \hline
\end{tabular}}
\end{table}

\begin{table}[H]
\caption{Image Augmentation and Vision Models Comparison}
\label{tab:Image_Model_Augmentation}
\begin{tabular}{|l|l|l|l|}
\hline
\textbf{Model} & \textbf{Augmentation} & \textbf{Epochs} & \textbf{Accuracy (\%)} \\ \hline
     Resnet50          &          No             &          49       &           70.3             \\ \hline
     Resnet50          &         Standard              &          80       &           76.1             \\ \hline
     Resnet50          &        RandomCrop + RandomRotation               &     177            &            81.7            \\ \hline
     Effnet-B3          &      RandomCrop + RandomRotation                 &      192           &       81.6                 \\ \hline
     VGG-11          &       RandomCrop + RandomRotation                &         491        &      78.1                  \\ \hline
      ViT         &         RandomCrop + RandomRotation              &        155         &           82.2             \\ \hline
     Baseline          &        Standard               &         N/A        &          84              \\ \hline
\end{tabular}
\end{table}

\begin{table}[H]
\caption{Accuracy Comparison Between Initial and Final Dataset}
\label{tab:Image_Model_Dataset}
\begin{tabular}{|l|l|l|}
\hline
\textbf{Model} & \textbf{Initial Dataset (\%)} & \textbf{Final Dataset (\%)} \\ \hline
      VGG-11         &                 78.1              &         12.7                    \\ \hline
      VGG-16         &                               &                12.7             \\ \hline
      Resnet-18         &                               &               66.6              \\ \hline
      Resnet-50         &              81.1                 &            69.9                 \\ \hline
      Resnet-152         &                               &               69.5              \\ \hline
      EfficientNet-0         &                               &             66.4                \\ \hline
      EfficientNet-3         &           81.6                    &            61.7                 \\ \hline
      EfficientNet-6         &                               &         21.5                    \\ \hline
      ViT         &          82.2                     &               21.5              \\ \hline
\end{tabular}
\end{table}

\begin{table}[H]
\caption{Transfer Learned Model vs. From Scratch}
\label{tab:Image_Model_Transfer}
\begin{tabular}{|l|ll|ll|}
\hline
\multirow{2}{*}{} & \multicolumn{2}{l|}{\textbf{Not Pretrained}} & \multicolumn{2}{l|}{\textbf{Transfer Learning}} \\ \cline{2-5} 
                  & \multicolumn{1}{l|}{Epochs}  & Accuracy (\%) & \multicolumn{1}{l|}{Epochs}   & Accuracy (\%)   \\ \hline
          VGG-11        & \multicolumn{1}{l|}{5}        &         12.7      & \multicolumn{1}{l|}{15}         &         33        \\ \hline
          VGG-16        & \multicolumn{1}{l|}{11}        &         12.7      & \multicolumn{1}{l|}{2}         &           29.7      \\ \hline
          Resnet-18        & \multicolumn{1}{l|}{150}        &       66.6        & \multicolumn{1}{l|}{49}         &          45.6       \\ \hline
          Resnet-50        & \multicolumn{1}{l|}{348}        &         69.9      & \multicolumn{1}{l|}{95}         &            60     \\ \hline
          Resnet-152        & \multicolumn{1}{l|}{212}        &       69.5        & \multicolumn{1}{l|}{61}         &           59.8      \\ \hline
          Efficientnet-0        & \multicolumn{1}{l|}{189}        &        66.4       & \multicolumn{1}{l|}{66}         &          53.7       \\ \hline
          Efficientnet-3        & \multicolumn{1}{l|}{167}        &      61.7         & \multicolumn{1}{l|}{118}         &        51.1         \\ \hline
          Efficientnet-6        & \multicolumn{1}{l|}{31}        &      21.5         & \multicolumn{1}{l|}{92}         &         43.7        \\ \hline
          ViT        & \multicolumn{1}{l|}{2}        &      21.5         & \multicolumn{1}{l|}{74}         &       64.4          \\ \hline
\end{tabular}
\end{table}

\begin{table}[H]
\caption{Effect of Image Resolution on Accuracy}
\label{tab:Image_Model_Resolution}
\begin{tabular}{|l|l|l|l|l|}
\hline
Model & Image Size        & Top-1 Accuracy (\%) & Top-3 Accuracy (\%) & Top-5 Accuracy (\%) \\ \hline
  Efficientnet0-75    & \multirow{3}{*}{224x224} &      72.8               &           88.9          &         93.4            \\ \cline{1-1} \cline{3-5} 
  ViT-75    &                   &      75.5               &        88.7             &      93               \\ \cline{1-1} \cline{3-5} 
  VGG11-75    &                   &           44.6          &        63.6             &     71.8                \\ \hline
 Baseline     & \multirow{4}{*}{300x300} &         72.1            &         86.5            &          91.1           \\ \cline{1-1} \cline{3-5} 
 \textbf{Resnet50-75}     &                   &         \textbf{80.1}            &      \textbf{92.1}              &              \textbf{95.2}       \\ \cline{1-1} \cline{3-5} 
  Efficientnet3-75    &                   &      76               &       89.9              &     93.8                \\ \cline{1-1} \cline{3-5} 
  Efficientnet0-75    &                   &        76.9             &        90.2             &        93.9             \\ \hline
  Resnet50-75    & \multirow{2}{*}{528x528} &          80.2           &       92              &          94.9           \\ \cline{1-1} \cline{3-5} 
  Efficientnet6-75    &                   &        76             &       90.1              &            93.8         \\ \hline
\end{tabular}
\end{table}

\begin{table}[H]
\caption{Baseline Text Classification}
\label{tab:Text_Model_Base}
\begin{tabular}{|l|l|l|}
\hline
Model                    & Augmentation   & Accuracy \\ \hline
\multirow{3}{*}{Mistral} & N/A            & 74.4     \\ \cline{2-3} 
                         & Choice Mapping & 76.9     \\ \cline{2-3} 
                         & Options        & 83.3     \\ \hline
\multirow{3}{*}{Llama}   & N/A            & 75.6     \\ \cline{2-3} 
                         & Choice Mapping & 88.5     \\ \cline{2-3} 
                         & Options        & 89.7     \\ \hline
\multirow{3}{*}{Falcon}  & N/A            & 66.7     \\ \cline{2-3} 
                         & Choice Mapping & 76.9     \\ \cline{2-3} 
                         & Options        & 38.5     \\ \hline
\end{tabular}
\end{table}

\begin{table}[H]
\caption{Augmentation Combinations}
\label{tab:Text_Model_Augment}
\begin{tabular}{|l|l|}
\hline
Prediction Amount + Choice Mapping/ Options & Accuracy (\%) \\ \hline
N/A                                         & 75.6          \\ \hline
Prediction-1                                & 96.2          \\ \hline
Prediction-3                                & 97.4          \\ \hline
Prediction-5                                & 93.6          \\ \hline
Choice Mapping                              & 88.5          \\ \hline
Prediction-1 + Choice Mapping               & 94.9          \\ \hline
Prediction-3 + Choice Mapping               & 93.6          \\ \hline
Prediction-5 + Choice Mapping               & 88.5          \\ \hline
Options                                     & 89.7          \\ \hline
Prediction-1 + Options                      & 94.9          \\ \hline
Prediction-3 + Options                      & 94.9          \\ \hline
Predictions-5 + Options                     & 81            \\ \hline
\end{tabular}
\end{table}

\begin{table}[H]
\caption{Prediction + Chain of Options}
\label{tab:Text_Model_Chain}
\resizebox{\columnwidth}{!}{\begin{tabular}{|l|l|l|l|}
\hline
                         &              & Accuracy- Predictions Only (\%) & Accuracy- Predictions + Chain of Options (\%) \\ \hline
\multirow{3}{*}{Llama}   & Prediction-1 & 97.2                            & 96.4                                          \\ \cline{2-4} 
                         & Prediction-3 & 95.9                            & 95.9                                          \\ \cline{2-4} 
                         & Prediction-5 & 93.3                            & 94.6                                          \\ \hline
\multirow{3}{*}{Mistral} & Prediction-1 & 94.9                            & 94.4                                          \\ \cline{2-4} 
                         & Prediction-3 & 91.3                            & 93.3                                          \\ \cline{2-4} 
                         & Prediction-5 & 90.0                            & 93.6                                          \\ \hline
\multirow{3}{*}{Falcon}  & Prediction-1 & 78.2                            & 90.5                                          \\ \cline{2-4} 
                         & Prediction-3 & 79.0                            & 90.0                                          \\ \cline{2-4} 
                         & Prediction-5 & 76.6                            & 94.6                                          \\ \hline
\end{tabular}}
\end{table}

\begin{table}[H]
\caption{Final Results of Resnet + Llama for Skin Disease Diagnosis}
\label{tab:Final_Model}
\begin{tabular}{|l|l|l|l|}
\hline
Model                           & Prediction & Training Method  & Accuracy (\%) \\ \hline
\multirow{6}{*}{Resnet + Llama} & 1          & Chain of Options & 86.3          \\ \cline{2-4} 
                                & 1          & Normal           & 90.1          \\ \cline{2-4} 
                                & 3          & Chain of Options & 88.3          \\ \cline{2-4} 
                                & 3          & Normal           & 90.5          \\ \cline{2-4} 
                                & \textbf{5}          & \textbf{Chain of Options} & \textbf{91.2}          \\ \cline{2-4} 
                                & 5          & Normal           & 86.5          \\ \hline
\end{tabular}
\end{table}

\section{Figures}

\begin{figure}[H]
\centering
\includegraphics[width=\linewidth]{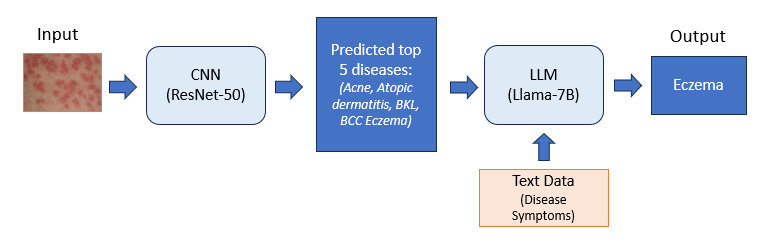}
\caption{Model Architecture for Skin Disease Classification}
\label{fig:attention_vis}
\end{figure}

\begin{figure}[H]
\centering
\includegraphics[width=\linewidth]{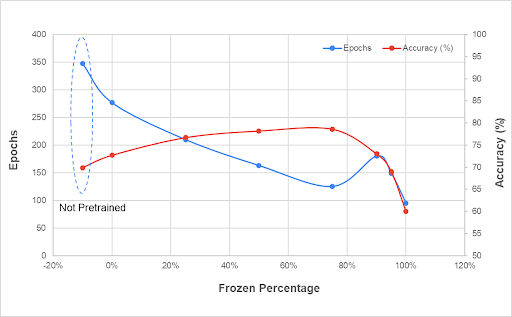}
\caption{Effect of pre-train frozen percentage on the accuracy for Resnet-50}
\label{fig:pre_train}
\end{figure}

\begin{figure}[H]
\centering
\includegraphics[width=\linewidth]{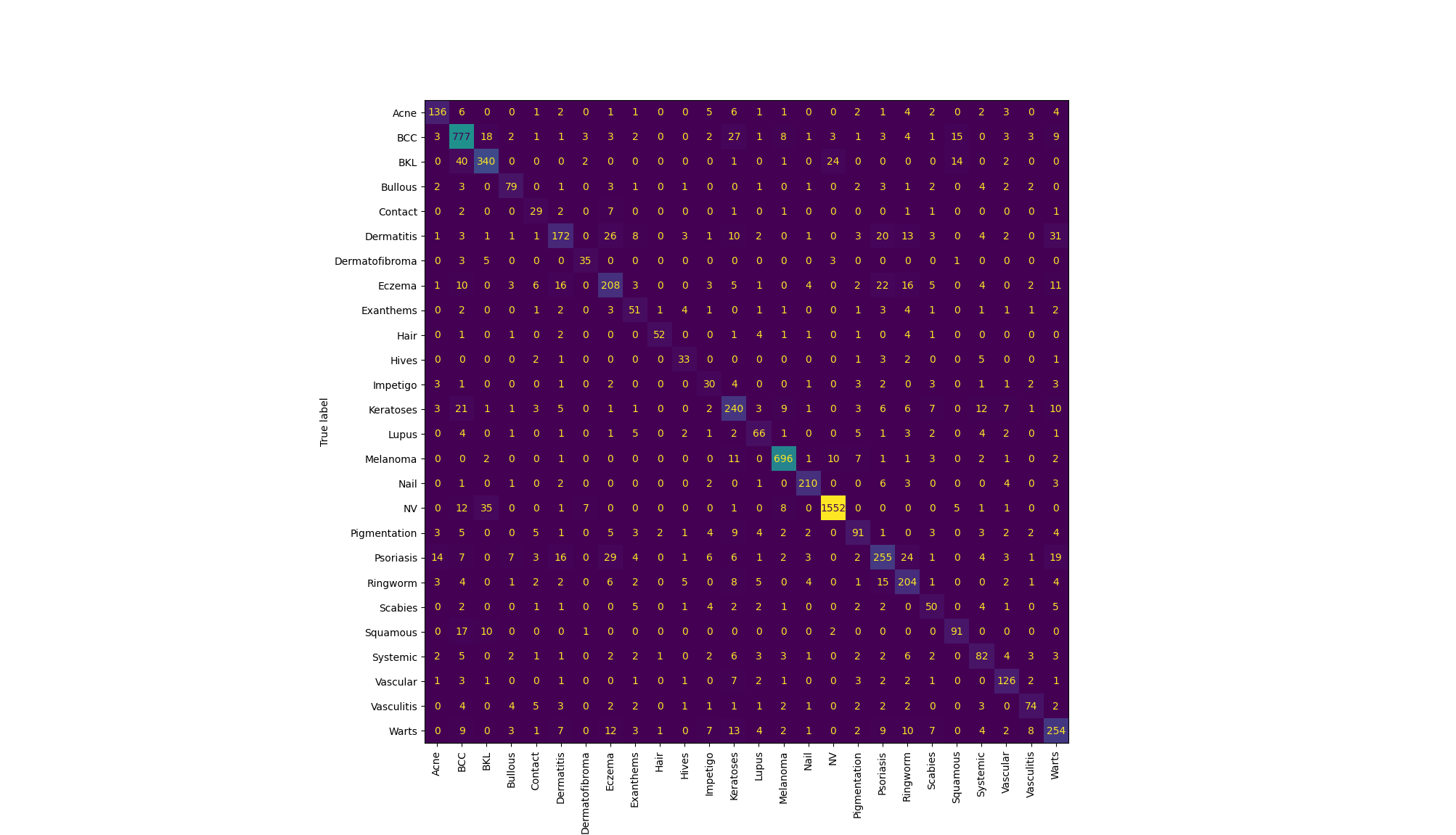}
\caption{Confusion Matrix of the Proposed Model (Resnet50) on the Final Dataset}
\label{fig:resnet}
\end{figure}

\begin{figure}[H]
\centering
\includegraphics[width=\linewidth]{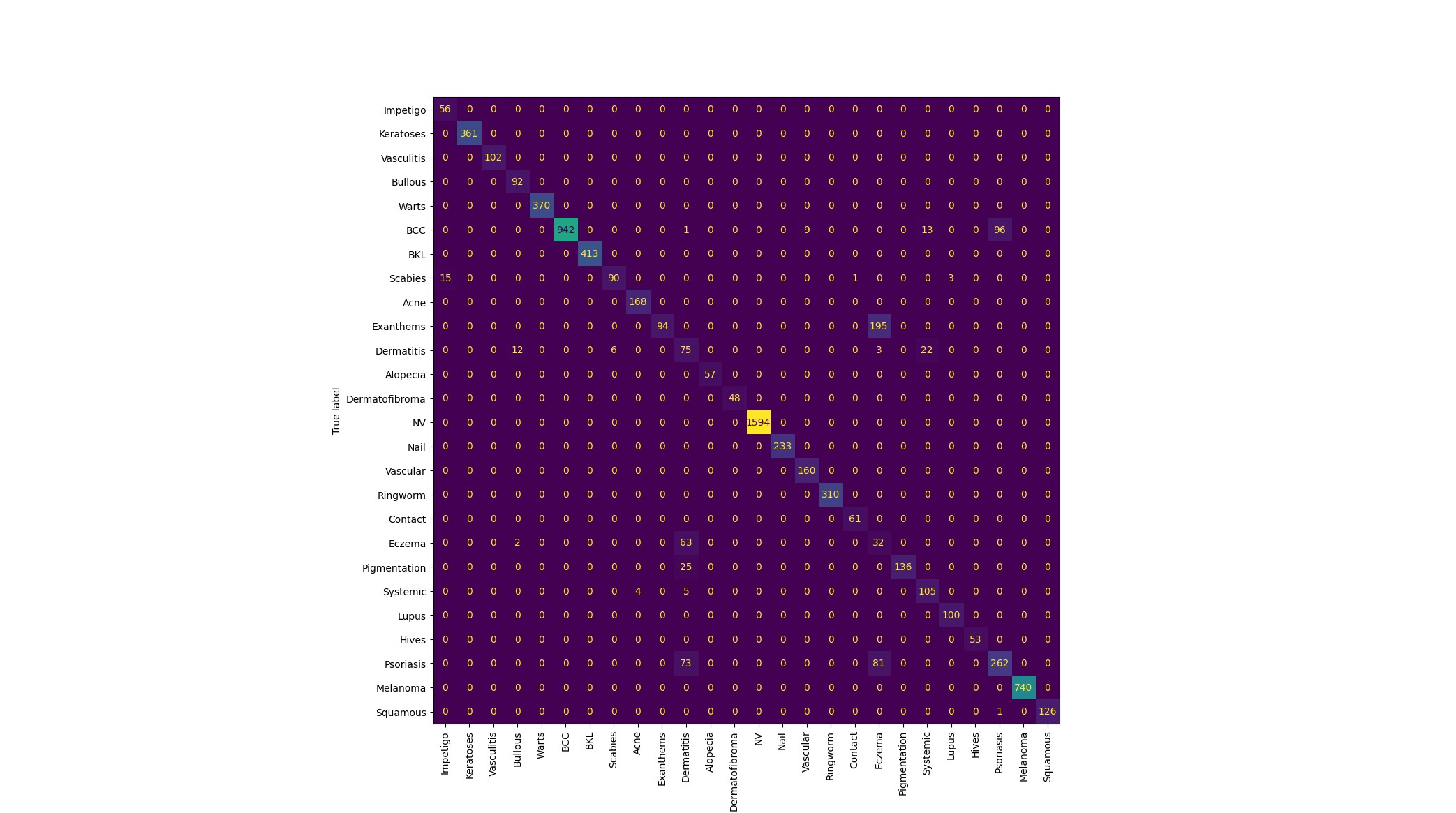}
\caption{Final Confusion Matrix with Image Count}
\label{fig:llm}
\end{figure}

\begin{figure}[H]
\centering
\includegraphics[width=\linewidth]{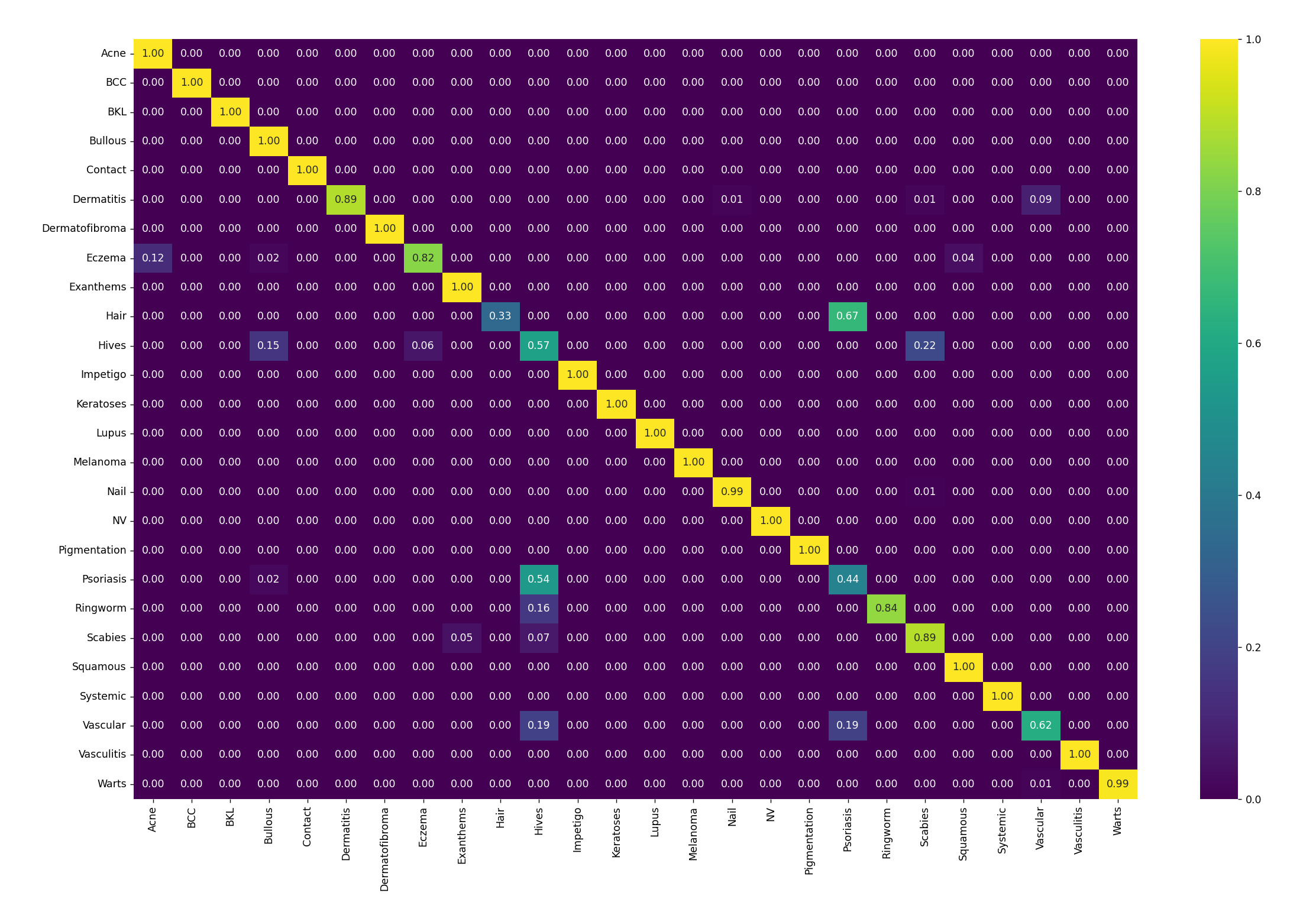}
\caption{Final Confusion Matrix with Accuracy}
\label{fig:llm_final}
\end{figure}

\end{document}